\documentclass[aps,twocolumn]{revtex4}
\begin{document}
\title{Comment on ``R$\ddot{o}$ntgen quantum phase shift: a semiclassical local electrodynamical effect?"}
\author{S.C. Tiwari}
\affiliation{Institute of Natural Philosophy, 1, Kusum Kutir, Mahamanapuri, Varanasi - 221 005, India}
\email{vns_sctiwari@yahoo.com}
\maketitle

The geometric phases in physics represent a large class of phenomena having experimental manifestation in some cases. 
The landmark papers that stimulated interest in this subject are those of Aharonov and Bohm \cite{1} and 
Berry \cite{2}. There has been an incessant debate on the origin of these phases: quantum or classical? 
Horsley and Babiker \cite{3} claim to have given a (semi-)classical derivation of what they term as 
`R$\ddot{o}$ntgen phase'. This phase could be viewed as dual to the Aharonov-Casher phase \cite{4} such 
that the motion of a neutral particle possessing an electric dipole moment is considered in the field of a 
magnetic line charge \cite{5}. In a nice analysis Wilkens \cite{5} shows that Aharanov-Bohm like 
topological phase arises due to the R$\ddot{o}$ntgen interaction term when the dipole moves in a force free region. 
In this comment I argue that the principal claim of \cite{3} is misleading.

The main contribution of \cite{3} is based on the following: 1) a cumulative impulse vector is introduced, 
2) for the radial force, Eq.(2) in \cite{3}, and the assumed circular motion of the dipole in the $xy$ plane, 
the $y$-component of the impulse vector is calculated, and 3) the difference in the displacement along the $y$
direction is identified with the phase shift. It is obvious that appearance of sin$\theta$ in the integrand 
for the impulse is responsible for the non-zero displacement along $y$ direction. Since this function i.e. 
sin$\theta$ would arise in any problem that has radial force, the electric dipole-magnetic line charge or 
magnetic dipole-electric charge construction is irrelevant, and the check on Aharonov-Casher phase is trivial. 
The authors do not give any physical justification or motivation for introducing the key element: cumulative
impulse vector. Note that in the classical collision problems, for instance, the instantaneous impulse forces 
are considered such that the instant force acts conserving the linear momentum. In the problem of dipole motion, 
if at all necessary, one may invoke impulse assuming that the force is suddenly switched on. But there would 
be no cumulative impulse vector. 

I think an important paper where the impulse approximation has been used on physical
grounds is that of electric charge and monopole scattering problem by Goldhaber \cite{6}. I have 
realized after reading this paper that the role of angular momentum in the geometric phases might
be of fundamental importance. Berry in his paper \cite{2} found a monopole of strength -1/2
located at the degeneracy for a typical Hamiltonian depending on three parameters. In an extensive
elaboration and beautiful analysis Aitchison \cite{7} discusses monopole-like structure of
strength - n/2, and interprets Dirac quantization condition with the spin of the system. In 
\cite{6}
spinless charge and spinless monopole scattering is considered, and the scattering angle is 
calculated for large impact parameter in the impulse approximation. A surprising result is
obtained, namely the azimuthal dependence of the scattering amplitude. An intriguing remark
is made by Goldhaber:" This brings out an important distinction between classical and quantum
theory. In a classical theory with an arbitrary force law there is no reason to expect a conserved
total angular momentum, even if energy and linear momentum are conserved. In quantum theory, 
general invariance requirements, combined with the linearity of the theory, guarantee the 
existence of a {\bf J} which commutes with the S matrix". Now the azimuthal dependence is compensated
introducing an extra factor $\chi$ ,and this is related with spin as well as Dirac quantization.
The spin is not an intrinsic property of monopole or electron, but depends on both. Is this
spin related with the geometric phase as conjectured in \cite{8} ? I think the difference
between classical and quantum description arises as we begin with the force laws such that in
the force-free and torque-free states the constant momentum and angular momentum states 
respectively are treated as equivalent. We have argued that Aharonov-Bohm effect and Berry phase
might be the manifestations of the inequivalence of these states \cite{9}. The tentative
suggestions may be of use to consider the scattering-oriented approach to the problem discussed
in \cite{3}

A simple illustrative example where time dependent angle occurs naturally is that of Foucault pendulum. A geometric phase, 
Hannay angle, is found in this case, see \cite{10}. Holstein has rightly remarked that the classical explanation given by
Boyer \cite{11} for the Aharonov-Casher phase has been disputed by Aharonov et al \cite{12}, ``asserting that any change in 
momentum associated with this `classical lag' effect goes into the internal momentum of the particle rather than into
modifying its kinetic momentum". The letter by Horsley and Babiker further neglects dipole-dipole interaction without 
making a relative estimate of all the forces that are to be included in principle. Though introduction of the Planck
constant 'by hand' does not necessarily imply quantum implication, I do believe that a semi-classical explanation
\cite{3} could offer useful insights.

In conclusion, the main claim of the authors in \cite{3}, namely the classical origin of the R$\ddot{o}$ntgen phase, 
is argued to be unfounded.

\end{document}